\documentclass[12pt,preprint]{aastex}

\shorttitle{Recurrent Novae Orbital Periods}
\shortauthors{Schaefer}

\begin{document}

\title{Orbital Periods For Three Recurrent Novae}

\author{Bradley E. Schaefer}
\affil{Physics and Astronomy, Louisiana State University,
    Baton Rouge, LA, 70803}

\begin{abstract}

I report on the discovery of the orbital periods of three recurrent novae in our galaxy.  V745 Sco has an orbital period of $510\pm20$ days with ellipsoidal modulations, based on SMARTS photometry from 2004-2008.  V3890 Sgr has an orbital period of $519.7\pm0.3$ days with ellipsoidal modulations and a shallow eclipse, based primarily on SMARTS and AAVSO photometry from 1995-2008, but also extending back to 1899 with archival plates.  In addition, a sinusoidal modulation of amplitude 0.2 mag and period $103.8\pm0.4$ days is seen mainly in the red, with this attributed to ordinary pulsations in the giant companion star.  V394 CrA has an orbital period equal to twice its primary photometric period ($P_{orb}=1.515682\pm0.000008$ days), as based on photometry extending from 1989-2008.  I use all available information (including the UBVRIJHK spectral energy distributions) to get distances to the four RNe with red giant companions as $800\pm140$ pc for T CrB, $4300\pm700$ pc for RS Oph, $7300\pm1200$ pc for V745 Sco, and $6000\pm1000$ pc for V3890 Sgr.  Further, the red giant in the RS Oph system has a mass loss rate of close to $3.7\times10^{-8}$ M$_{\odot}$ yr$^{-1}$ as based on many confident measures, and this is too weak (by a factor of 100,000) to supply the white dwarf with mass at the known rate of $3.9\times10^{-6}$ M$_{\odot}$ yr$^{-1}$.  Thus, the only way to get matter onto the white dwarf fast enough is through Roche lobe overflow, and this confidently demonstrates that the distance to RS Oph is $\gtrsim3000$ pc.

\end{abstract}

\section{Introduction}

	Recurrent novae (RNe) are binary systems that undergo ordinary nova eruptions with recurrence time scales less than a century or so (Webbink et al. 1987).  Systems can have a fast recurrence time only if their white dwarf is near the Chandrasekhar mass and if their accretion rate is very high (Prialnik \& Kovetz 1995).  With this required situation, RNe will likely have the white dwarf mass increasing until it collapses as a Type Ia supernova, and so RNe are a likely solution (at least in part) to the important and long-standing progenitor problem.  This problem has recently come to even higher prominence because the identity of the progenitor must be known for calculations of evolution effects to be made, as such effects can change the luminosity-decline relation as we look back to an older, lower-metallicity Universe, with associated systematic offsets in the supernova Hubble diagram (Dom\'inguez, H\"oflich, \& Straniero 2001).  But before we can conclude that RNe are Type Ia progenitors, we must determine whether there are enough RNe to produce the observed supernova rate and we must test whether the white dwarf loses more mass during each eruption than it gains by accretion between eruptions.  Both of these questions depend on RN demographics.
	
	With our discovery that V2487 Oph (Nova Oph 1998) had a `second' nova eruption in 1900 (Pagnotta, Schaefer, \& Xiao 2008), there are now ten known RNe in our own galaxy (Schaefer 2009).  The single most important parameter of any cataclysmic variable system is the orbital period, $P_{orb}$.  This is especially true for RNe, as their known periods range from 0.076 days for T Pyx (Schaefer et al. 1992) to 455.7 days for RS Oph (Fekel et al. 2000).  While almost all of the cataclysmic variables have $P_{orb}<0.3$ days, four of the six RNe have known periods $>0.6$ days.  And two of the systems with unknown periods (V745 Sco and V3890 Sgr) must be long because they have infrared excesses and spectra that indicate red giant companion stars (Harrison, Johnson, \& Spyromilio 1993).  
	
	With the most important demographic property for RNe being known for only six-out-of-ten RNe, I started a long running program to seek the orbital periods for the three systems V745 Sco, V3890 Sgr, and V394 CrA.  (V2487 Oph was discovered to be a RN too recently to be part of this program.)  The idea is to seek photometric modulations with the orbital period.  A difficulty of this method is that two of the targets apparently have long orbital periods, so many years are required to accumulate enough cycles to provide a convincing periodicity.  Other difficulties are that all cataclysmic variables display flickering and brightness changes on all time scales that lead to large scatter in any folded light curve, and that a photometric period might be half of the true orbital period due to ellipsoidal effects.

\section{Data}
	
	All magnitudes were taken by differential photometry with respect to nearby comparison stars.  These comparison stars were measured by relative photometry with respect to standard stars reported by Landolt (1992).  These calibrations were made several times for each field under photometric conditions with the SMARTS 1.3-meter telescope on Cerro Tololo.  The magnitude systems throughout are the Johnson U, B, and V bands and the Kron-Cousins R and I bands.  The measurement of the comparison star magnitudes in the infrared (J, H, and K bands) were taken directly from the 2MASS sky survey (Cutri et al. 2003).  The accuracy of the comparison star magnitudes is roughly 0.015 mag for an absolute calibration.
	
	The primary data source for my period-search program has been the SMARTS telescopes on Cerro Tololo in Chile, mainly the 1.3-m telescope but including the 0.9-m and 1.0-m telescopes also.  The specific program started in 2004, but I have additional data (on these same telescopes) for V394 CrA from many observing runs back to 1989.  In 2004, I also have concentrated observations of V745 Sco and V3890 Sgr on most of the nights throughout a 14-night observing run.  For V745 Sco, my data from SMARTS consist of 516 R-band magnitudes on 246 nights from June 2004 to August 2008, plus several runs of nearly-simultaneous BVRIJHK photometry.  For V3890 Sgr, my SMARTS data consist of 350 R-band magnitudes plus 105 J-band magnitudes on 141 nights from June 2004 to September 2008 plus several nearly-simultaneous runs of BVRIJHK photometry.  For V394 CrA, I have 237 magnitudes (almost all in the B-band) from week-long runs in 1989, 1994, 1995, 1996, and 1997, plus 263 magnitudes (mostly in B-band and R-band) well spread out in the five observing seasons from July 2004 to July 2008.
	
	For V3890 Sgr, I have also collected data from three additional sources over the last 109 years.  I have measured 54 B-band magnitudes from many archival plates at the Harvard College Observatory from 1899 to 1939.  The typical one-sigma uncertainty is 0.15 mag, and the magnitudes are on the modern B-band magnitude system.  I have measured 465 R-band magnitudes on 340 nights with the ROTSE IIIa telescope in Australia from 2003 to 2005.  The median uncertainty in the differential magnitude is 0.02 mag.  The CCD images were taken with no filter and are thus broad band, with the effective central wavelength for the resulting magnitudes being close to the I-band (as determined from comparing relative magnitudes for comparison stars of similar color).  Additionally, I collected 68 V-band magnitudes from the {\it American Association of Variable Star Observers} (AAVSO) database from 1995 to 2005.  The typical one-sigma uncertainties are 0.15 mag, and the magnitudes are in the Johnson V-band magnitude system.

\section{V745 Sco}

	V745 Sco had nova eruptions in 1937 and 1989, reaching a peak magnitude of V=9.4 mag and the time from the peak until the light curve had declined by three magnitudes ($t_3$) is 9 days.  After my modern photometry of comparison star magnitudes, remeasuring the outburst brightnesses on the original archival plates, and collecting all existing data, I conclude that the eruption light curve was identical for both events (Schaefer 2009).  In quiescence, I measured average magnitudes and colors to be R=15.5, B-V=1.9, V-R=2.6 mag.
	
	The intensive photometry during the two-week run in June 2004 shows that V745 Sco always displays the usual flickering with durations of 30-60 minutes and amplitudes 0.1-0.2 mag.  I do not have any time series in other than the R-band, so I cannot determine whether the flickering light is relatively blue or red in color. The light curve also displays secular trends with time scales from a day to a week to a year, with a total range of 2.5 mag.  Superposed on these trends are significant maxima and minima with typical time scales of $\sim 200$ days.
	
	I have made a discrete Fourier transform of my nightly-averaged light curve.  To avoid the effects of the long-term secular changes dominating the transform, I have subtracted out a fitted parabola, and I get the same results when I subtract out simple yearly averages.  There is only one prominent highly-significant peak (other than a lower peak at exactly half a year caused by my sampling structure), which has a period of $255\pm10$ days.  A cause for concern is that my data cover only six photometric periods, so it is possible that randomly scattered shot noise could reproduce the high Fourier peak.  Another question is whether the orbital period is double the photometric period.  With the strong precedent of T CrB (Zamanov et al. 2004), a sinusoidal modulation of this period and amplitude should be from ellipsoidal effects.  A strong observational reason for this doubling is that both the minima and the maxima for the folded light curve alternate between high and low (see Figure 1).  In particular, 14 of 16 observations with magnitude brighter than 15.1 are close to the 0.25 phase maximum, while all 16 observations with magnitude fainter than 15.9 are close to the 0.0 phase minimum.  This significant difference between alternating maxima and minima demonstrates that the photometric modulation is not caused by red giant oscillations.  This shape is typical for other RNe (specifically, CI Aql, U Sco, V394 CrA, and V3890 Sgr) as based on my own SMARTS light curves.  As such, I take the orbital period to be twice the photometric period, so $P_{orb}=510\pm20$ days.  The zero phase (taken as the time when the red giant is in conjunction in front of the white dwarf) is the Julian date of the deepest minimum is $2453800+N\times 510$.
  
\section{V3890 Sgr}

	V3890 Sgr had nova eruptions in 1962 and 1990, reaching a peak magnitude of V=8.1 mag with $t_3=14$ days.  Again, with detailed measures of the original plates and comparison stars as well as a collection of all existing eruption data, I find that the two eruptions had identical light curves (Schaefer 2009).  In quiescence, I measure the average brightness and colors to be R=14.3, B-V=0.9, and V-R=1.2 mag.  Harrison, Johnson, \& Spryomilio (1993), Gonzalez-Riestra (1992), and Anupama \& Sethi (1994) have measured the extinction, I have also estimated an extinction from the light curve colors, and the median of all the measures is $E_{B-V}=0.9$ mag.  Harrison, Johnson, \& Spyromilio (1993) gave a spectral classification of the companion star as M5 III.
	
	Long runs in June 2004 showed continuous flickering with time scales of tens of minutes and amplitudes of $\sim 0.03$ mag, as well as secular trends lasting$>10$ days.  The amplitude of this flickering increases greatly toward the blue (see next paragraph).  The long-term variations are generally from 14.0-14.5 in the R-band, although I have seen excursions to 13.7 and 15.1 mag.
	
	The SMARTS J-band data and the ROTSE I-band data both show a highly significant periodicity at $103.8\pm0.4$ days.  In the infrared, the folded light curve shows a good sine wave with full amplitude of 0.14 mag and a small scatter of 0.04 mag.  The amplitude of this signal does not change greatly with wavelength, but the scatter (due to flickering and other variations) increases greatly towards the blue (being 0.04, 0.15, 0.23, 0.44, and 0.57 mag in the J, I, R, V, and B bands respectively).  The spectral energy distribution of V3890 Sgr (see Section 6) shows that the J-band flux is dominated by the red giant by more than a factor of a thousand, so the 103.8 day periodicity must be tied to the  red giant.  Low amplitude variations with typical periods of 100 days are the norm for red giants (Fraser, Hawley, \& Cook 2008), so I attribute the 103.8 day period to small pulsations on the red giant.
	
	The SMARTS R-band data show a single prominent periodicity at a period of $257\pm3$ days, while the AAVSO V-band data also have a prominent peak in the Fourier transform at $265\pm3$ days.  The Harvard B-band data have a moderately significant peak in its Fourier transform at $258.2\pm0.8$ days.  A Fourier transform of all three data sets (after normalizing the average of each data set) has a single highly significant peak at a period of $259.85\pm0.15$ days. The amplitude of this periodic component increases steadily towards the blue, being roughly 0.07, $\lesssim$0.2, 0.5, 0.7, and 1.0 for the J, I, R, V, and B bands respectively.  The star gets bluer in color as it brightens, with the V-R color going from 1.5 to 1.2 as its V magnitude brightens from 16.1 to 15.3 mag.  The stability of this period from 1899 to 2008 shows that it is tied to the orbital period.  Again, with the strong precedent of T CrB, we have to expect that the true orbital period might be double the photometric period.  The folded light curve on the double period (Figure 2) shows a pronounced difference between the depths of alternating minima, apparently with a shallow eclipse superposed on ellipsoidal variations plus ordinary flickering.  With this, the true orbital period is $519.7\pm0.3$ days and the ephemeris for the primary minimum is Julian date $2454730+N\times519.7$.

\section{V394 CrA}

	V394 CrA had nova eruptions in 1949 and 1987, with a peak V-band magnitude of 7.2 mag and $t_3=5.2$ days.  Again, I have collected all the existing eruption magnitudes, measured the original archival plates, and placed the magnitudes onto a modern magnitude scale, with the result being that a single light curve template fits both eruptions well (Schaefer 2009).  In quiescence, I measure an average magnitude of R=17.9 mag, while the colors are B-V=0.7 and V-R=0.5 mag.  Duerbeck (1988) examined nearby field stars and concluded that $E_{B-V}\sim 0.25$ mag for all distances past one kiloparsec, however the colors during eruption are much too blue ($B-V=-0.2$ mag at peak) even for zero extinction.  Schaefer (1990) found a highly significant photometric periodicity of 0.7577 days (with amplitude 0.6 mag) from one observing run in 1989.
	
	Follow-up photometry with the Cerro Tololo 0.9-m telescope during four observing runs in 1994, 1995, 1996, and 1997 found V394 CrA to be brighter than in 1989 (by one magnitude on average) and to have a smaller amplitude ($\sim 0.3$ mag).  V394 CrA displays the usual flickering on time scales from minutes to hours with amplitudes up to half a magnitude.  I do not have quasi-simultaneous light curves in multiple bands, so I have not measured the color relative color of the flickering light.  Nevertheless, the star appears to get bluer in color as it gets brighter.  On long time scales, the star varies widely, from 17.0 to 19.3 mag in the R-band (with B-band data converted to R-band using the average color).  V394 CrA continued the pattern of having relatively small amplitude when it was bright (such as in 2008) and a large amplitude when it was faint (as in 2005).  Detailed calculations demonstrate that this behavior is consistent with a slow-changing light source (presumably from variable accretion) superposed on an unchanging periodic signal (presumably from ellipsoidal and possible eclipse effects).
	
	I have performed discrete Fourier transforms on my light curves, where the yearly averages (in each color) have been subtracted out.  The 2005 data have a single highly-significant peak at $0.7580\pm 0.0003$ days.  Taken together, the 1989 and 2005 data have too long a gap to keep a good cycle count, but this can be resolved with the intervening data.    By far the best alias (with a Fourier power of 50.3 times the average power, while the second highest peak is only 40.2 times the average) is with a period of $0.757841\pm0.000004$ days.  The stability of the photometric periodicity from 1989 to 2005 demonstrates that it must be tied to the orbital period.
	
	With the strong precedents of U Sco (in red bands) and CI Aql (as based on my extensive SMARTS light curves, see for example Hachisu, Kato, \& Schaefer 2003), we have to consider that this photometric period is only half the orbital period.  This question can be resolved by finding whether the odd-numbered cycles behave differently from the even-numbered cycles.  That is, in a light curve folded on a period of 1.515682 days, are the two maxima (at phases 0.25 and 0.75) of different height (like for CI Aql), and are the two minima (at phases 0.0 and 0.5) of different depth (like for U Sco)?  Figure 3 shows the V394 CrA light curve folded on twice the photometric period (for the 1989 and 2005 data alone), and we see apparent differences, with the primary minimum (around zero phase) having almost all the magnitudes that are fainter than 19.43 (even going to 20.02 mag) while the secondary minimum never gets fainter than 19.43 mag.  For averages in phase bins of size 0.2, the two maxima are at $18.78\pm0.02$ and $18.92\pm0.05$.  For averages in phase bins of size 0.1, the primary minimum has an average of $19.25\pm0.07$, while the secondary minimum has an average of $19.13\pm0.02$.  Based on the precedent of CI Aql, the maximal asymmetry is between the differences from each minimum to the following maximum.  For V394 CrA, the primary-to-following-maximum amplitude is $0.47\pm0.07$ mag, the secondary-to-following-maximum amplitude is $0.21\pm0.05$ mag, with these being different by $0.26\pm0.08$ mag which is at the 3.0-$\sigma$ confidence level.  As such, I conclude that the orbital period is $1.515682\pm0.000008$ days.  The Julian dates for the primary minima are given as $2453660.81+N\times 1.515682$.

\section{RN Distances}

	The second most important parameter for RN demographics is the distance.  Here, I will calculate the distances to the four RNe with red giant companions based on three methods.  The input, intermediate values, and derived distances are presented in Table 1.
	
	The first two methods calculate a distance to the red giant assuming that it fills its Roche lobe of radius ($R_{Roche}$) derived from its orbital period ($P_{orb}$).  (This assumption is demonstrated to be correct in the next Section.)  The binary separation is $a=4.16R_{\odot}(M_{wd}+M_{comp})^{1/3}P_{orb}^{2/3}$ where the period is expressed in days, and the Roche lobe size is $R_{Roche}=0.49~a~q^{2/3}/[0.6q^{2/3}+ln(1+q^{1/3})]$ where the mass ratio is $q=M_{comp}/M_{wd}$ (Frank, King, \& Raine 2002).  I take the mass of the white dwarf mass ($M_{wd}$) to be 1.35 $M_{\odot}$ and red giant mass ($M_{comp}$) to be 1.0 $M_{\odot}$.  These masses are reasonable (cf. Hachisu \& Kato 2001), and the quoted size has only weak dependence on this assumption.   The extinction ($E_{B-V}$) is taken as the median of reported values and from the light curve colors.  
	
	The first method uses my K-band magnitudes ($m_K$), which are corrected to bolometric magnitudes as $m_{bol}=m_K-0.33*E_{B-V}+(V-K)+BC$, with $(V-K)+BC$ equaling 2.62 to 2.86 mag for M0 to M6 (Johnson 1966).  The spectral types give $T_{eff}$, the red giant radius is taken to be the Roche lobe size, and the absolute bolometric magnitude is $M_{bol}=42.36-10\log(T_{eff})-5\log(R_{Roche}/R_{\odot})$.  The distance in parsecs is $D=10^{(m_{bol}-M_{bol}+5)/5}$.  The second follows the same path (with the same $R_{Roche}$ and $E_{B-V}$) but uses different means to get temperatures and brightnesses.  My derived extinction-corrected spectral energy distributions (see Figure 4 and Schaefer 1986) give the frequency of maximum flux ($\nu_{max}$), the flux at that frequency ($f_{\nu 0}(\nu_{max})$), and the resulting effective temperature ($T_{eff}$).  For blackbody emission, the temperature give the surface flux ($F_{\nu}(\nu_{max})=T_{eff}^3\times5.96 \times 10^{-16}$ erg cm$^{-2}$ s$^{-1}$ Hz$^{-1}$), and the inverse square law then gives the distance ($D=R_{Roche}(F_{\nu}/f_{\nu 0})^{1/2}$).  A comparison between my derived distances for the two methods shows fractional deviations from the averages of 13\%, 26\%, 7\%, and 22\% (with an average of 17\%).  So I take the uncertainty arising from the various input measures to be 17\%.
	
	A third method is based on the outburst light curve.  The `maximum magnitude versus rate of decline' (MMRD) relations connect the decline time over the first two and three magnitudes from peak ($t_2$ and $t_3$) to the absolute visual magnitude as $M_V=-10.79+1.53\log(t_2)$ and $M_V=-11.26+1.58\log(t_3)$ as appropriate for fast events (Downes \& Duerbeck 2000).  Another relation is that $M_V=-5.23$ at 15 days after the peak (van den Bergh \& Younger 1987).  From these three relations I get three measures of $M_V$ (which have RMS scatters of two-thirds of a magnitude) and the distance moduli $\mu$, which I average to get the best value.  The distance is  $D=10^{(\mu-3.1E_{B-V}+5)/5}$.  The scatter and calibration issues are large for this third method, so the results from the first two methods are strongly preferred.
	
	Many RN distances have been published, but the ones based on the red giant and on the MMRD are now superseded by those in Table 1.  The only remaining independent published distances are for RS Oph (see the next Section), all of which have critical and questionable assumptions plus large uncertainties.  In all, the best distance is just the average of the two distances based on the red giant along with the 17\% uncertainty.

\section{Does the Red Giant in RS Oph Fill Its Roche Lobe?}
	
	A substantial question is whether the red giant in RS Oph fills its Roche lobe.  If the red giant fills its Roche lobe then Roche lobe overflow provides a simple means of getting matter onto the white dwarf, whereas if the red giant does not fill its Roche lobe then the only way to supply matter onto the white dwarf is through accretion from a wind emitted by the companion star.  
	
	The question of Roche lobe filling is equivalent to asking the distance to RS Oph.  That is, with analyses similar to that in the last section, we can determine the size of the red giant as a function of its distance, and by knowing the size of the Roche lobe (from the orbital period) we can determine whether the red giant is filling it.  The results from the previous section demonstrate that the red giant would be close to filling its Roche lobe for any distance $\gtrsim$3000 pc, with various uncertainties making the precise limit uncertain, although the results from the previous section can be expressed as $4300\pm700$ pc.

\subsection{Previous Estimates of the Distance to RS Oph}
	
	Previous modelers have taken the red giant size to be a fraction of $R_{Roche}$, based entirely on an adopted small distance scale.  For example, Hachisu \& Kato (2001) and Dobrzycka et al. (1996) adopted distances of 600 pc and 1500 pc so as to conclude that the red giant filled 25\% and 60\% of its Roche lobe respectively.  These small presumed distances were based on the selection of some distance from amongst the many previous estimates.  With the model results being based on assumption, we can only look to the observational evidence as to the distance to RS Oph.  In this subsection, I will examine the many published distances for RS Oph.
	
	The MMRD relations have been frequently used to get a distance to RS Oph, primarily because they can be made with only a light curve.  Typical values for three MMRD relations are presented in Table 1 as method 3, with an average derived distance of 2100 pc.  Unfortunately, the uncertainties in this method are very large.  Downes \& Duerbeck (2000) find a one-sigma scatter about the best fit MMRD relation to be 0.6 mag with maximum deviation of 1.6 mag, with this corresponding to errors in distance by 30\% up to over a factor of two.  Even the different versions of the MMRD fitted to the same data differ by up to 1.0 mag (a factor of 1.6 error).  For the case of RNe, the MMRD has given distances for U Sco of up to 95,000 pc (Webbink 1978), while V2487 Oph has a similarly implausible MMRD distance far outside our galaxy.  On the basis of this evidence, errors of over a factor of two are common for MMRD distance estimates as applicable to RNe.  The plausible range of MMRD distances can be expressed as something like 1050-4200 pc.
	
	In the 2006 eruption, an infrared source visible only by means of interferometry was interpreted as being the companion star next to the bright nova envelope, with the angular separation (taken to be the binary separation) then giving an upper limit on the distance of $\lesssim$540 pc (Monnier et al. 2006).  But this distance measure has the critical assumption that one of the infrared sources is the red giant star.  The source is $\sim10\times$ brighter than the red giant in quiescence.  The authors present the binary hypothesis only as an idea, while presenting alternative ideas that the infrared asymmetry is caused by a circumbinary reservoir of heated gas (much like for T Pyx, Shara et al. 1997), an expanding wind, or a jet.  Indeed, Lane et al. (2007) explain the infrared shell as simply the expected light from the asymmetric expanding nova envelope.  With this work by Lane et al., the weak and tentative distance estimate of Monnier et al. is based on a misinterpretation of the infrared extension and so no distance estimate can be made based on the observations.  In all, the confidence of the binary explanation is too weak to support a distance estimate of any useful reliability.
	
	Rupen, Mioduszewski, \& Sokoloski (2008) detected an asymmetric synchrotron shell during the 2006 eruption and inferred an expansion parallax of $2450\pm400$ pc.  The big trouble with the distance is that we have no real way of knowing which {\it radial} velocity selected from an optical emission line profile corresponds to the {\it transverse} expansion velocity appropriate for some selected contour level in a highly asymmetric radio shell.  The critical assumption is that the half-width at zero intensity (HWZI) of the velocity along the line of sight corresponds to the part where the outer edge of the  inner shell of radio brightness has fallen to near one-third of its peak.  But HWZI values are always hard to measure against a variable continuum, and there is no reason to choose the HWZI instead of the HWHM (or any other position in the profile of the emission lines).  In addition, we have to assume that a highly asymmetric shell (claimed to be bipolar) has its {\it transverse} velocity in one direction equal to the {\it radial} velocity towards Earth.  Also, the radio observations show {\it two} radio shells, so the quoted distance estimate has to assume that the relevant expansion is that of the {\it inner} radio shell and not the {\it outer} radio shell.  Finally and critically, the position of the edge of the radio shell is arbitrary, as for example O'Brien et al. (2006) chose the peak brightness radius of the inner shell which is over 20\% smaller, while another reasonable choice would be to take the outermost radius of the inner shell with detected radio flux.  Thus, the quoted distance requires three assumptions for each of which a range of a factor of two is plausible.  The error bars quoted by Rupen, Mioduszewski, \& Sokoloski (2008) include only their measurement uncertainty, whereas the systematic uncertainty is likely more like a factor of two or larger.  Their expansion parallax does provide real information about the distance to RS Oph, but with a more realistic total uncertainty of a factor of two, the constraint on the distance is 1200-4900 pc.
	
	Hjellming et al. (1986) measured the column density for HI and deduced a distance of 1600 pc. This is based on an assumed {\it linear} relation between column density and distance, which might be valid on average for the plane, but for which the patchiness is notorious and so the derived distances have large uncertainties.  Unfortunately, the quoted equation is not mentioned in the quoted citation.  But the real problem is that the 1600 pc distance is taken from a linear relation which does not attempt to account for the fall off of hydrogen density with height above the plane.  RS Oph has a galactic latitude of 19.5$\degr$, so the line of sight quickly passes above the region for which the equation might be applicable.  Hence, this commonly quoted distance to RS Oph cannot be used.  The hydrogen column density does carry distance information, but only as a lower limit.  This limit can only be something like $>$1000 pc for when the line of sight passes out of the disk.  
	
	Cassatella et al. (1985) provide another frequently quoted distance claim that RS Oph must be closer than the Carina Arm of our galaxy ($\sim$2000 pc distant) because no extinction component is seen with a velocity of +20 km s$^{-1}$, despite the difficulty to separate this component from the local arm.  But again, RS Oph has a galactic latitude of 19.5$\degr$, so our line of sight passes over the Carina Arm at a height of 670 pc.  With the gas in the disk having a Gaussian scale height of 135 pc, we realize that there will be no absorption from the Carina Arm for any distance to RS Oph.  (That is, the gas density at 670 pc from the midplane is roughly $4\times10^{-6}$ times the density at midplane, with this being completely unobservable.)  Thus, the `Carina Arm limit' is not valid for RS Oph.
	
	Livio, Truran, \& Webbink (1986) used the measured brightness, temperature, and orbital period of the red giant to derive a blackbody distance to RS Oph of 3180-3290 pc.  They are using effectively method 1 from Table 1, and they get an identical answer even though their input for masses and period differ somewhat from the values in the Table.  The critical assumption for this distance estimate is whether the red giant fills its Roche lobe.  Up to now in this paper, we have been undecided (but see the next two sections for a decisive answer), so this previous distance estimate might expressed as `3200?' pc.
	
	Let me now summarize the realistic distance estimates to RS Oph based on previously published evidence.  We have 1050-4200 pc, no limit, 1200-4900 pc, $>$1000 pc, no limit, and 3200? pc, respectively for the paragraphs above.  All the prior distance estimates have substantial uncertainties and critical assumptions, so we must try to minimize the problems.  We see the center of the indicated ranges is $\sim$3000 pc, which would point to the red giant filling its Roche lobe.  But the allowed range from these prior published evidences is too large, and it spans the distances for which RS Oph must or must-not fill its Roche lobe.  So I conclude that the prior evidences are not helpful for deciding the question of whether the red giant fills its Roche lobe.

\subsection{Wind Accretion Model}

	Some modelers of the RS Oph system (e.g., Hachisu \& Kato 2001; Dobrzycka et al. 1996) have assumed that the red giant does not fill its Roche lobe, and they provide the material onto the white dwarf by a wind coming from the red giant.  The idea is that red giants normally emit a wind, and some fraction of this will accrete onto the white dwarf to provide the material needed for the nova eruption.  If the red giant does not fill its Roche lobe, then the wind accretion idea is the only mechanism to provide material to the white dwarf.
	
	A critical problem with any model without Roche lobe filling is that the required red giant wind is greatly too weak to supply the white dwarf at the known rate.  That is, we know the mass loss rate of the red giant from multiple independent methods, we know what fraction of this mass can be accreted onto the white dwarf, and we know how much mass is actually accreted onto the white dwarf from multiple independent methods.  The result is that wind accretion fails by orders of magnitude to provide enough mass to the white dwarf.  I will now provide the logic and details of this analysis.  
	
	First, we know the mass loss rate in the red giant wind from several methods.  First, a red giant for the known stellar parameters should have a loss rate of  $\dot{M}_{wind}=3.7\times10^{-8}$ M$_{\odot}$ yr$^{-1}$ (Kudritzki \& Reimers 1978).  Second, detailed modeling of the {\it IRAS} and {\it Spitzer} infrared spectra gives $2.3\times10^{-8}$ M$_{\odot}$ yr$^{-1}$ (Evans 1987; van Loon 2008).  Third, detailed modeling of the x-ray emission during the 2006 eruption gives $1.4\times10^{-7}$ M$_{\odot}$ yr$^{-1}$ (Vaytet, O'Brien, \& Bode 2007).  Fourth, the lack of any radio detection forces the wind to be $<3\times10^{-7}$ M$_{\odot}$ yr$^{-1}$ (Seaquist \& Taylor 1990).  In all, the wind strength is close to $3.7\times10^{-8}$ M$_{\odot}$ yr$^{-1}$ and certainly in the range $2 - 30 \times10^{-8}$ M$_{\odot}$ yr$^{-1}$.  
	
	Second, the red giant wind velocity has been directly measured to be 90 km s$^{-1}$ from the observed widths of narrow coronal emission lines produced when the 1985 eruption ionized the wind (Shore et al. 1996).  This velocity would correspond to the velocity along the line of sight to Earth.  With RS Oph having a fairly low inclination (Dobrzycka \& Kenyon 1994), this means that we are looking roughly along the direction perpendicular to the orbital plane.  Walder et al. (2008) calculates in a 3D simulation of wind accretion in the RS Oph system that the wind velocity in the orbital plane is 2.5 times larger than in the polewards direction.  So we would expect a wind velocity of over 200 km s$^{-1}$ in the direction towards the white dwarf, with this being the relevant quantity for considering accretion onto the white dwarf.  With this, the ratio of the wind velocity to the orbital velocity (`R'  in the notation of Walder et al., 2008, and `$V_R/A\Omega$' in the notation of Nagae et al. 2004) is 5.4.  Thus, the accretion onto the white dwarf is firmly in the regime of Bondi accretion (Walder et al. 2008; Nagae et al. 2004).
	
	Third, we know the fraction of the red giant wind that will get accreted onto the white dwarf from independent calculations.  Livio, Truran, \& Webbink (1986) calculate that only 1/170 of the wind will get onto the white dwarf of the RS Oph system for Bondi accretion.  This was calculated for a wind velocity equal to the escape velocity of the red giant of 100 km $^{-1}$.  Nagae et al. (2004) perform a detailed hydrodynamical calculation and find that the accretion is a factor of five smaller than the simple Bondi accretion value.  With this, the Livio, Truran, \& Webbink (1986) result should be reduced to something like 0.001.  With the better wind velocity (as described in item 2), the fraction of the red giant wind accreted onto the white dwarf will be 0.0001-0.0002 (Nagae et al. 2004).  Walder et al. (2008) also performed a hydrodynamical calculation for the accretion fraction, but they presumed an unrealistically low wind velocity of 20 km s$^{-1}$, and this error leads to an overestimate of the accretion fraction by 2-3 orders of magnitude.  So the best estimate of the fraction of the wind falling onto the white dwarf is something like 0.001 (corresponding to the relevant wind velocity being 100 km $^{-1}$), while the plausible range might be from 0.0001-0.005.  
	
	Fourth, items one and three can be combined to estimate the mass accretion rate onto the white dwarf in the case of wind accretion.  The best estimate is  $3.7\times10^{-11}$ M$_{\odot}$ yr$^{-1}$ while the value is certainly less than  $1.5\times10^{-9}$ M$_{\odot}$ yr$^{-1}$.  
	
	Fifth, we also know how much mass is required to accumulate on the white dwarf between eruptions from two separate methods.  Livio, Truran, \& Webbink (1986) calculate the critical accumulated mass to be $5.9\times10^{-5}$ M$_{\odot}$ for the expected case, and they can push it to an extreme of $8.9\times10^{-7}$ M$_{\odot}$.  Detailed modeling of the x-ray production during the 2006 nova eruption gives the mass of the ejecta as $1.1\times10^{-6}$ M$_{\odot}$ (Vaytet, O'Brien, \& Bode 2007), so the accreted mass can only be larger.  Hachisu \& Kato (2001) make detailed models of the optical light curve during eruption to calculate that the ejected envelope has a mass of $2\times10^{-6}$ M$_{\odot}$, and again the accreted mass can only be larger.  The best estimate of the accumulated mass is $5.9\times10^{-5}$ M$_{\odot}$, while it is certainly greater than $8.9\times10^{-7}$ M$_{\odot}$.  
	
	Sixth, since 1890, RS Oph has had eight nova eruptions in 1898, 1907, 1933, 1945, 1958, 1967, 1985, and 2006 (Schaefer 2009).  With this, the average recurrence time scale is $(2008-1890)/8=15$ years.  
	
	Seventh, items five and six can be combined to give the required average accretion rate onto the white dwarf required for an eruption.  The best estimate is $3.9\times10^{-6}$ M$_{\odot}$ yr$^{-1}$, while the extreme limit is $5.9\times10^{-8}$ M$_{\odot}$ yr$^{-1}$.  
	
	Eighth, we can now compare the required accretion rate (item seven) against what a red giant wind can provide (item four).  For the best estimate values, the red giant can supply mass only at the rate of $3.7\times10^{-11}$ M$_{\odot}$ yr$^{-1}$, while we know that the white dwarf is being fed mass at a rate of $3.9\times10^{-6}$ M$_{\odot}$ yr$^{-1}$.  That is, the red giant wind can only provide $10^{-5}$ of what is known to get onto the white dwarf.  Even by simultaneously pushing everything to its extreme limits, the red giant wind still fails by a factor of 40.  The best estimate of the red giant mass loss (item one) is already less than the required accretion rate onto the white dwarf (item seven) by a factor of 100, so even pushing to 100\% accretion fraction cannot solve the problem by two orders of magnitude.
	
	Ninth, a strong conclusion can now be made, based on this simple comparison with the input reliably known from multiple independent methods.  The red giant wind is too weak by a factor of something like 100,000$\times$ to be able to provide the required mass to the white dwarf.  This signals the complete failure of the wind accretion model.

\subsection{Roche Lobe Overflow Model}

	The wind accretion model fails by many orders of magnitude to provide material to the white dwarf at the observed rate.  But the material has to get there.  With no prospect for a solution from the red giant wind, the only other possibility is by Roche lobe overflow.  This is the default scenario, as all other classical novae and RNe are powered by Roche lobe overflow.  With Roche lobe overflow being the universal case amongst fast novae and it being the only possibility, we have a simple and strong case that RS Oph has its red giant filling its Roche lobe.
	
	Nevertheless, the Roche lobe overflow model might have problems.  We have to make some overall evaluation of any problems with the Roche lobe overflow model as compared to the severe problem of the wind accretion model.  We need to reach some conclusion on the distance scale (and the model for the mass transfer) that minimizes the problems.  In this subsection, I will examine various potential worries and problems.
	
	A potential worry is that a distance of 4300 pc will make the peak absolute magnitude for RS Oph equal to $M_V=m_V - 5\log D +5-3.1E_{B-V}$ or $M_V$ equals -10.6 mag.  Actually, with the measurement uncertainties between the two methods (see Table 1), we only need a distance of  $\gtrsim$3000 pc for Roche lobe filling, so the limit on $M_V$ is only -9.8 mag.  This is indeed substantially more luminous than most classical novae, but such a comparison is fatally flawed because most novae are greatly slower than RS Oph so by the MMRD relations they will systematically be much less luminous.  Even with this highly biased comparison, RS Oph has a lower peak luminosity than 14\% of novae with expansion parallaxes (Shafter 1997) and 14\% of the novae used to calibrate the MMRD relations (Downes \& Duerbeck 2000).  Indeed, two novae (CP Pup and V1500 Cyg) have $M_V$ equal to -10.7 mag (Downes \& Duerbeck 2000).  The absolute magnitude of RS Oph is fully consistent with the MMRD relations, which predict $M_V=-9.5$ with a scatter of 0.6 mag.  In all, there are good precedents for a very fast nova like RS Oph to have the absolute magnitude required for the case where the red giant fills its Roche lobe.
	
	A theoretical version of this same worry is that the bright absolute magnitude for RS Oph (for the distance associated with Roche lobe filling) is too bright to be possible in theoretical models.  Indeed, typical nova models lead to peak absolute magnitudes around -8 (Prialnik \& Kovetz 1995).  Yet, especially luminous peaks can come from either of two conditions, high CNO abundances (Starrfield et al. 2000) or fast eruptions on massive white dwarfs (Prialnik \& Kovetz 1995).  RS Oph has already been recognized as having ``strong evidence" for substantial enhancements in the CNO elements as based on both ultraviolet absorption lines (Shore et al. 1996) and infrared spectroscopy (Evans et al. 1988).  And RS Oph is a very fast nova (with $t_3=14$ days) and the white dwarf must be near the Chandrasekhar mass (e.g., Hachisu \& Kato 2001).  With RS Oph satisfying both conditions for high peak luminosity (CNO enhancements and a fast eruption on a high mass white dwarf), the theoretical version of the worry about the peak absolute magnitude goes away.
	
	Another potential problem can arise if the Roche lobe overflow from the red giant becomes dynamically unstable leading to a very high accretion rate.  Indeed, ``A dynamical mass transfer instability can occur when mass is being transferred from the more massive to the less massive component and the mass donor has a deep convective envelope, as in the case of a giant or asymptotic giant branch (AGB) star losing mass to a less massive companion." (Rasio \& Livio 1996).  If this case were realized in the RS Oph system, then as soon as the Roche lobe is filled, the mass transfer would increase on a dynamical time scale with accretion rates reaching $\sim10^{-4}$ M$_{\odot}$ yr$^{-1}$ and a common envelope would be formed.  However, this case {\it requires} that the red giant be more massive than the white dwarf, or actually, that the mass ratio (for the red giant to the white dwarf) be $>5/6$ (Frank, King, \& Raine 2002).  For the RS Oph system, the red giant is 0.5-1.2 M$_{\odot}$ (Hachisu \& Kato 2001), or 0.5 M$_{\odot}$ (Dobrzycka \& Kenyon 1994; Shore et al. 1996).  There is an inclination dependence for the red giant mass, varying from around 0.5-1.2 M$_{\odot}$ over the preferred range of inclination as determined from the orbit and light curve (Dobrzycka \& Kenyon 1994).   The white dwarf mass is certainly near the Chandrasekhar limit (so that the recurrence time scale can be so fast) and is near 1.35-1.377 M$_{\odot}$ (Hachisu \& Kato 2001).  And just recently, Brandi et al. (2009) have reported a radial velocity curve for the white dwarf and the red giant, and thus they measure a mass ratio of $0.59\pm0.05$.  With this, we see that the RS Oph system is far from the unstable accretion case, and so the potential problem of runaway instabilities in the accretion is not applicable.

	If the red giant fills its Roche lobe, then the RS Oph system might display ellipsoidal variations, with such never having been reported in the literature (Oppenheimer \& Mattei 1993; Dobrzycka \& Kenyon 1994).  (Actually, the red giant should also display ellipsoidal variations if it does {\it not} fill its Roche lobe, so the lack of these variations is also a potential worry, albeit a lesser worry, for the wind accretion models.)  However, there are two good reasons to know why RS Oph should {\it not} show ellipsoidal modulations that are independent of the question for whether it is filling its Roche lobe.  First, the orbital inclination is likely to be small, and the amplitude of the ellipsoidal variations will be small in such a case.  Dobrzycka \& Kenyon (1994) perform an extensive analysis of the radial velocity and light curve data to conclude that that inclination is $30\degr-40\degr$.  Second, RS Oph has large chaotic variations on all time scales (from minutes to decades) with amplitudes of up to 2.5 magnitudes (based on the AAVSO light curve since 1934).  With these fast and large amplitude variations, the expected ellipsoidal effects will be lost in the noise.  As such, the lack of apparent ellipsoidal modulations is not a real worry for the Roche lobe overflow models.
	
	In all, we are left with no problems with the Roche lobe overflow models.  And we still know that Roche lobe overflow is the only way to get matter onto the white dwarf at the required rate and that Roche lobe overflow is a universal property of all novae.  So we have the strong conclusion that the red giant in the RS Oph system is filling its Roche lobe, and hence that the distance to RS Oph must be $\gtrsim$3000 pc and something like $4300\pm700$ pc.

\section{Conclusions}

	This paper presents orbital periods for three of the ten galactic RNe, and fairly accurate distances for all four with red giant companions.  I also demonstrate that the red giant in the RS Oph system must be filling its Roche lobe, because Roche lobe overflow is the only way to feed the white dwarf at the high observed rate.
	
	The obvious next step in the investigation of these systems is to get radial velocity curves.  This will serve two purposes.  First, a period from the radial velocities can serve to confirm the periods presented in this paper.  While simple in principle, this will not be easy in practice.  For the case of V394 CrA, the closest analogous system is U Sco (with a similar period, spectrum, and quiescent magnitude), yet even the confirmation of its eclipse period is fraught with problems and errors (Johnston \& Kulkarni 1992; Schaefer \& Ringwald 1995).  For the systems with red giant companions, the necessity of getting a well-sampled radial velocity curve over multiple orbits will require a program involving dozens of observing runs spread over many years.  And the small orbital velocities will require good spectral resolution beyond that which can be obtained by common low-resolution spectrometers.
	
	Second, radial velocity curves for these systems can provide a measure of the white dwarf mass.  Yet for this, we have the long-standing problem that the velocities of the emission lines (from the accretion disk) generally do not track the velocity of the white dwarf.  This often leads to unreliable white dwarf masses (cf. Wade 1985; Robinson 1992; Schaefer \& Ringwald 1995).  In the case of RNe, we have the poor track record that the majority of the mass determinations are far from the Chandrasekhar mass (Johnston \& Kulkarni 1992; Kraft 1958).  It might be possible to use some observing technique to pull out the white dwarf velocity (Bitner \& Robinson 2006; Thoroughgood et al. 2001), although the reliability of these methods will have to be demonstrated.  The goal of this is to test the strong theoretical claim that the mass of the white dwarfs in RNe must be close to the Chandrasekhar mass.

\clearpage

\begin{deluxetable}{lllll}
\tabletypesize{\scriptsize}
\tablecaption{All RN Distance Information
\label{tbl1}}
\tablewidth{0pt}
\tablehead{
\colhead{~}   &
\colhead{T CrB}   &
\colhead{RS Oph}   &
\colhead{V745 Sco}  &
\colhead{V3890 Sgr}
}
\startdata

$P_{orb}$ (d)	&	227.57	&	453.6	&	510	&	519.7	\\
$a$ ($R_{\odot}$)	&	208	&	331	&	357	&	362	\\
$R_{Roche}$ ($R_{\odot}$)	&	74	&	117	&	126	&	128	\\
$E_{B-V}$ (mag)	&	0.1	&	0.73	&	1.0	&	0.9	\\
Method 1:	&		&		&		&		\\
~~$m_K$ (mag)	&	4.8	&	6.5	&	8.3	&	8.3	\\
~~$(V-K)+BC$ (mag)	&	2.80	&	2.62	&	2.86	&	2.83	\\
~~Spectral Type	&	M4 III	&	M0 III	&	M6 III	&	M5 III	\\
~~$T_{eff}$ (K)	&	3400	&	3700	&	3300	&	3400	\\
~~$m_{bol}$ (mag)	&	7.57	&	8.88	&	10.83	&	10.83	\\
~~$M_{bol}$ (mag)	&	-2.29	&	-3.66	&	-3.33	&	-3.49	\\
~~D (pc)	&	900	&	3200	&	6800	&	7300	\\
Method 2:	&		&		&		&		\\
~~$\log [\nu_{max}]$ (Hz)	&	14.25	&	14.32	&	14.32	&	14.27	\\
~~$f_{\nu 0}(\nu_{max})$ (erg cm$^{-2}$ s$^{-1}$ Hz$^{-1}$)	&	$1.0\times10^{-22}$	&	$6.8\times10^{-24}$	&	$3.7\times10^{-24}$	&	$7.2\times10^{-24}$	\\
~~$T_{eff}$ (K)	&	3000	&	3600	&	3600	&	3200	\\
~~D (pc)	&	700	&	5400	&	7800	&	4700	\\
Method 3:	&		&		&		&		\\
~~$m_V$ at peak (mag)	&	2.5	&	4.8	&	9.4	&	8.1	\\
~~$m_V$ at 15d (mag)	&	8.4	&	7.9	&	13.2	&	11.2	\\
~~$t_2$ (days)	&	4.0	&	6.8	&	6.2	&	6.4	\\
~~$t_3$ (days)	&	6.0	&	14.0	&	9.0	&	14.4	\\
~~Average $\mu$ (mag)	&	12.84	&	13.90	&	18.85	&	17.21	\\
~~D (pc)	&	3200	&	2100	&	14100	&	7600	\\
Conclusions:	&		&		&		&		\\
~~D (pc)	&	$800\pm140$	&	$4300\pm700$	&	$7300\pm1200$	&	$6000\pm1000$	\\

\enddata
    
\end{deluxetable}

\clearpage
\begin{figure}
\epsscale{1.0}
\plotone{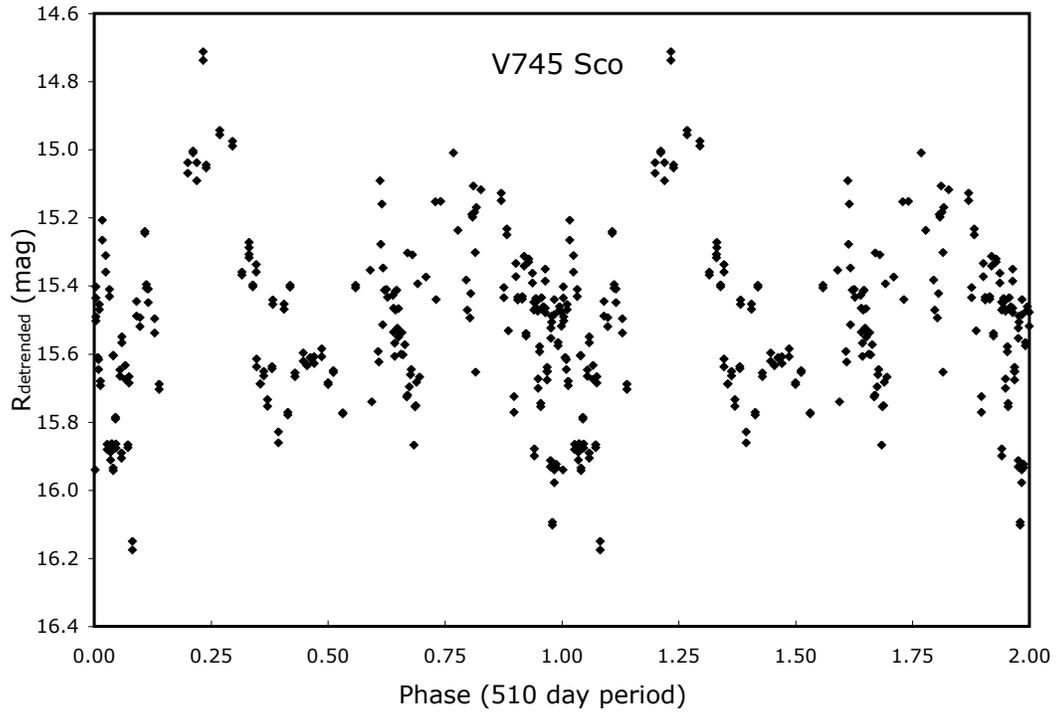}
\caption{
Folded light curve for V745 Sco.  The SMARTS 2004-2008 data are shown as nightly averages on the orbital period of 510 days, with the phase being doubled and the data repeated.  The usual flickering is superposed on a sinusoidal photometric modulation that is characteristic of ellipsoidal effects on the red giant companion star.
}
\end{figure}

\clearpage
\begin{figure}
\epsscale{1.0}
\plotone{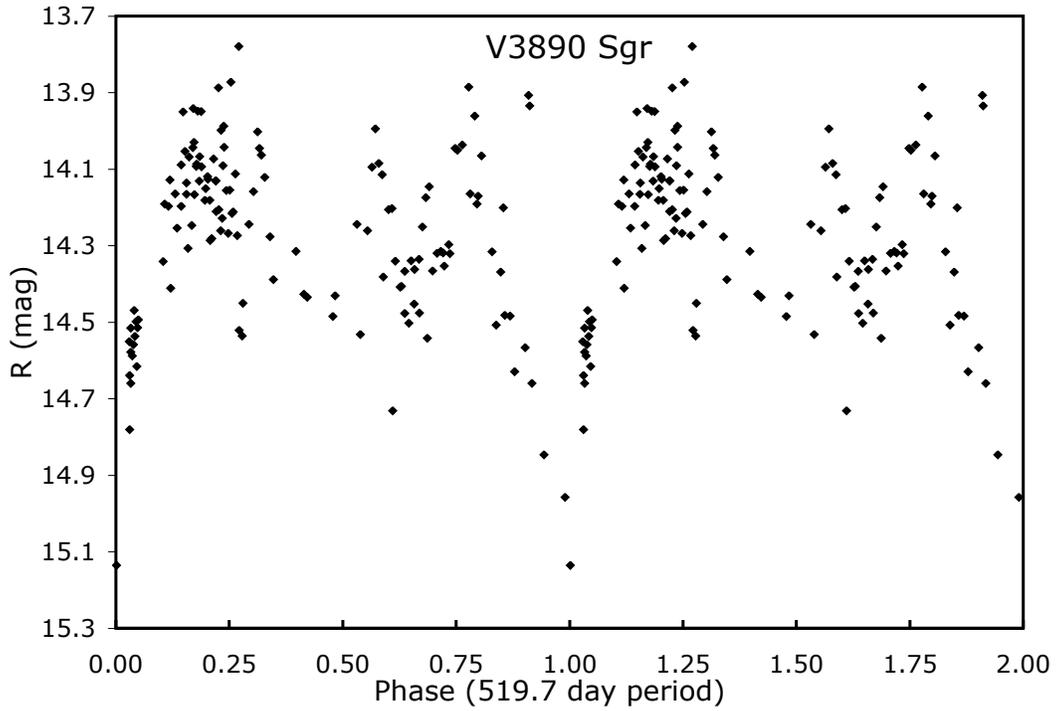}
\caption{
Folded light curve for V3890 Sgr.  The SMARTS 2004-2008 data are shown as nightly averages on the photometric period of 519.7 days, with the phase being doubled and the data repeated.  We see the usual flickering superposed on a sinusoidal photometric modulation that is characteristic of ellipsoidal effects on the red giant companion star plus what might be a shallow eclipse.
}
\end{figure}

\clearpage
\begin{figure}
\epsscale{1.0}
\plotone{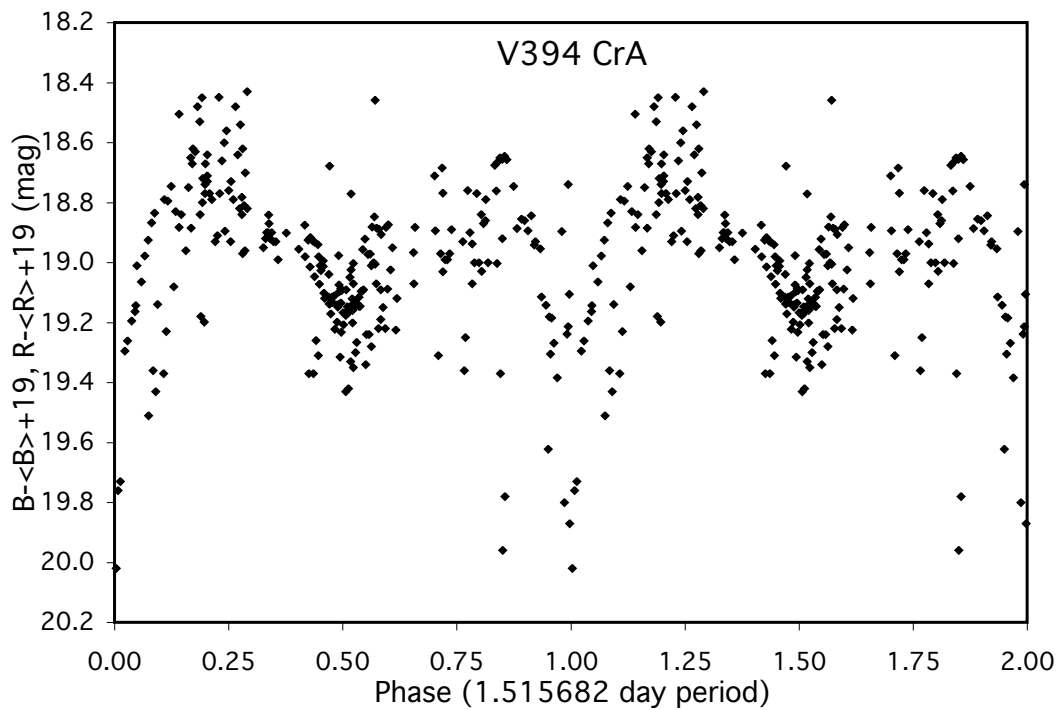}
\caption{
Folded light curve for V394 CrA.  The SMARTS 2004-2008 data are shown (with B and R data normalized together by the subtraction of yearly means) on the photometric period of 1.515682 days.  The primary minimum is substantially deeper than the secondary minimum, while the two maxima are also of different heights.  As with the similar system CI Aql, the overall modulation is likely from ellipsoidal effects on the secondary star, the unequal-maxima are caused by a asymmetric emission of the hot spot, and the unequal-minima are cause by a partial eclipse of the accretion disk.
}
\end{figure}

\clearpage
\begin{figure}
\epsscale{1.0}
\plotone{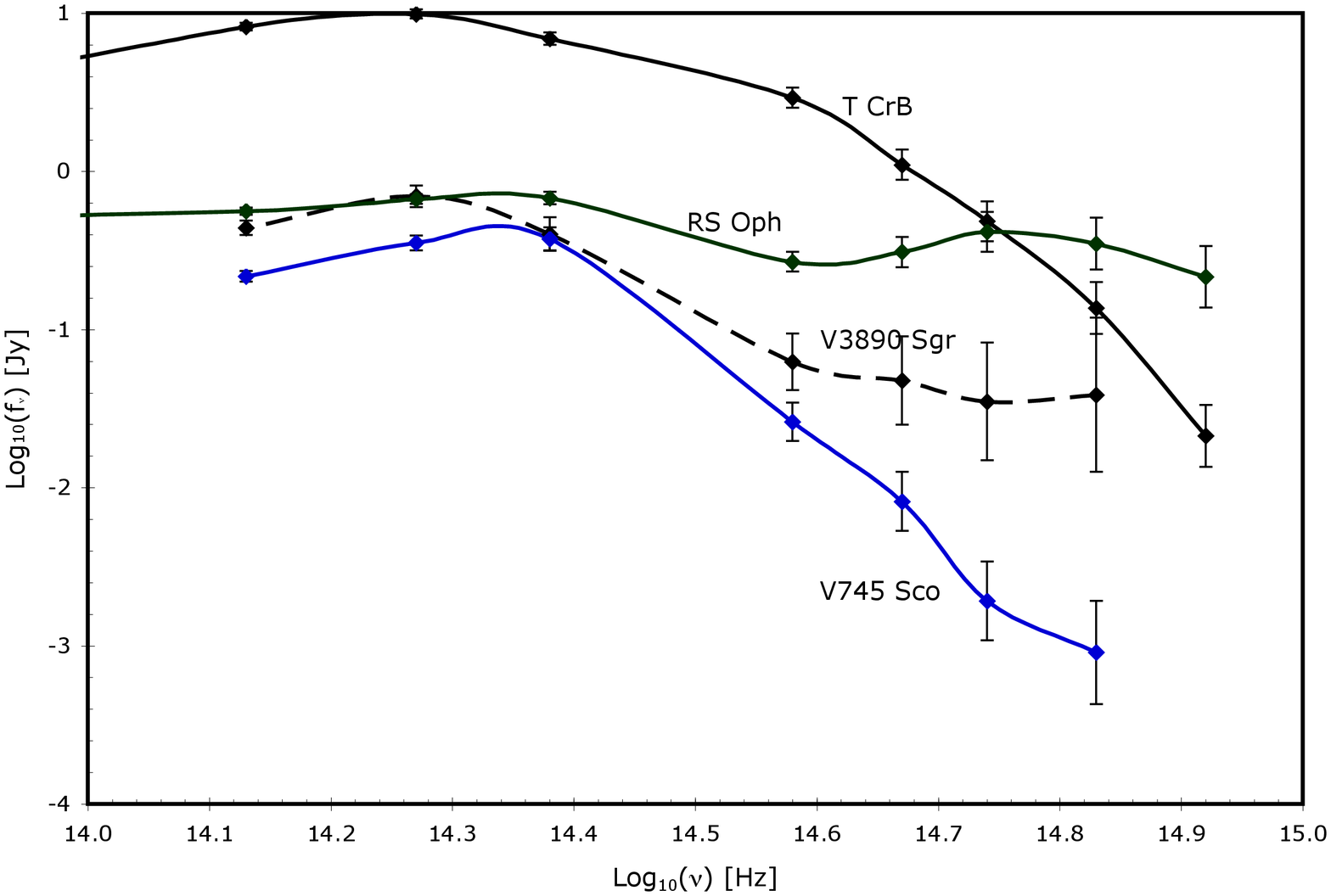}
\caption{
Spectral energy distributions for RNe with red giant companions.  All four have infrared excess associated with the red giant and perhaps some other sources.  The second distance method described in this paper assumes that the peak gives the temperature and flux from the red giant.  Any contribution from another source will only push the derived distances further away.
}
\end{figure}

\end{document}